\documentclass[aps,prb,twocolumn,amsmath,amssymb,superscriptaddress,floatfix]{revtex4}
\usepackage{graphicx}
\usepackage{bm}
\bibstyle{apsrev.bib}
\def\opsim{\mathop{\sim}}

\newcommand{\be}{\begin{equation}}
\newcommand{\ee}{\end{equation}}
\newcommand{\beqn}{\begin{eqnarray}}
\newcommand{\eeqn}{\end{eqnarray}}

\begin{document}

\title{Griffiths-McCoy singularities in random quantum spin chains: Exact results}

\author{Ferenc Igl\'oi}
\email{igloi@szfki.hu}
\affiliation{Research Institute for Solid State Physics and Optics,
H-1525 Budapest, P.O.Box 49, Hungary}
 \affiliation{Institute of Theoretical Physics,
Szeged University, H-6720 Szeged, Hungary}
\author{Istv\'an A. Kov\'acs}
\email{steve3281@bolyai.elte.hu}
\affiliation{Department of Physics, Lor\'and E\"otv\"os University, H-1117 Budapest,
P\'azm\'any P. s. 1/A, Hungary}
\affiliation{Research Institute for Solid State Physics and Optics,
H-1525 Budapest, P.O.Box 49, Hungary}
\date{\today}

\begin{abstract}
We consider random quantum ({\it tight-binding, XX} and {\it Ising}) spin chains
in the off-critical region and study their Griffiths-McCoy singularities.
These are obtained from the density of states of the low-energy excitations, which is
calculated exactly by the Dyson-Schmidt method. In large finite systems the low-energy excitations
are shown to follow the statistics of extremes and their distribution is given by the Fr\'echet form.
Relation between the Dyson-Schmidt technique and the strong disorder renormalization group
method is also discussed.

\end{abstract}


\maketitle

\section{Introduction}
\label{sec:intro}

Quenched disorder has a profound effect on the low-energy, low-temperature and long
wavelength properties of quantum systems. The interplay between quantum fluctuations,
correlations and disorder fluctuations generally results in strong singularities in the
thermodynamical quantities and in the (dynamical) correlation functions\cite{qsg,im}. This type of effect
takes place even outside the quantum critical region, e.g. in the quantum
paramagnetic phase at zero temperature, $T=0$, where spatial correlations are short ranged\cite{im,vojta}. The origin of this phenomenon, as pointed
out by Griffiths\cite{griffiths}, is due to rare regions, in which
strong bonds are accumulated by extreme fluctuations, so that the system in these
regions is locally in the thermodynamically unstable ferromagnetic phase. As a consequence
the excitation energy, $E$, in the rare
regions is very small, the relaxation process is very slow and the associated relaxation time,
$\tau \sim E^{-1}$, is divergent in the
thermodynamic limit. If we consider a finite part of a sample with linear size, $\ell$, the characteristic
time-scale of the slowest relaxation process stays also finite and asymptotically given by:
\be
\tau \sim \ell^{z}\;.
\label{tau}
\ee
Here $z=z(\delta)$ is the dynamical exponent, which is generally a continuous function of the
quantum control parameter, $\delta$, which measures the distance from the quantum critical point.

According to scaling theory\cite{scaling,ijr} the distribution of the low-energy excitations, $n(E,\ell)$, depends
on the scaling combination, $E\ell^z$, and for a small but fixed $E$ it is proportional to the volume, $\ell^d$, since the probability of finding a rare region goes linearly with the volume. From this
the asymptotic behavior of the distribution function in the thermodynamic limit reads as:
\be
n(E) \sim E^{d/z-1}\;.
\label{n_E}
\ee
Thermodynamical quantities which are obtained through integration of the density of states
are also singular.
For example the low-temperature
behavior of the average linear susceptibility, $\chi(T)$, and that of the specific heat, $c_v(T)$,
is expected to scale as\cite{im,vojta}
\be
\chi(T)\sim T^{-1+d/z},\quad c_v(T)\sim T^{d/z}\;,
\label{chi_cv}
\ee
whereas the small-field, $H$, dependence of the zero-temperature magnetization is given by:
\be
m(H) \sim H^{d/z}\;.
\label{m_H}
\ee
One can see from Eq.(\ref{chi_cv}) that the susceptibility is divergent at zero temperature for $z(\delta)>d$,
which was noticed first by McCoy\cite{mccoy} in an exact calculation of the random transverse-field Ising chain (RTFIC).

Detailed results about Griffiths-McCoy singularities are obtained for
one-dimensional (1d) systems partially by numerical investigations
(free-fermionic techniques\cite{free_fermion,bigpaper,ijr},
density matrix renormalization method\cite{DMRG,ijl},
quantum Monte Carlo (MC) simulations\cite{QMC}) and by analytical
calculations\cite{fisher,s1,ijl,i02} based on
the use of a strong disorder renormalization group (SDRG) method\cite{im}.
For higher dimensional systems Griffiths-McCoy singularities are studied numerically, either
by quantum MC simulations\cite{QMC1} or by numerical implementation of the SDRG method\cite{SDRG1}.

Analytical and conjecturedly exact results about Griffiths-McCoy singularities are scarce and
these are practically restricted to the RTFIC. Analytical solution of the SDRG equations
is first obtained in the vicinity of
the quantum critical point\cite{fisher}, i.e. in the weakly disordered
and in the weakly ordered Griffiths phases, where the dynamical exponent is shown to diverge as
$z(\delta) \sim 1/|\delta|$. The solution is then extended to the complete Griffiths phase\cite{ijl,i02}
and the calculated value of $z(\delta)$ is shown to agree with that obtained through a mapping
to a random walk problem in a random environment\cite{ir98}.

In this paper we use a direct and simple method to calculate exact values of the Griffiths-McCoy
singularities in a class of random quantum spin chains. These models include the random tight-binding
chain, the random antiferromagnetic XX-chain as well as the RTFIC. The low-energy
excitations for each models have the same form: they are obtained from the eigenvalue problem
of a symmetric tridiagonal matrix, $\cal M$ see in Eq.(\ref{M_tb}), with random (positive) entries. In the off-critical region of the spin chains there is an even-odd asymmetry: the matrix-elements of $\cal M$
are taken from different
distributions at even and odd bonds. We calculate the density of states, $n(E)$, in the
center of the band by the Dyson-Schmidt technique\cite{dyson} using the random walk
idea by Eggarter and Riedinger\cite{eggarter_riedinger}.
In Ref.[\onlinecite{eggarter_riedinger}] $n(E)$ is calculated in the continuum approximation
for an even-odd symmetric $\cal M$, which corresponds to the critical point of the random quantum spin chains.
In the present paper $\cal M$ has a general even-odd asymmetric form, which
corresponds to a strongly disordered quantum Griffiths phase and for which the continuum approximation
is no longer valid. Having the exact behavior of $n(E)$ at hand we than calculate the singularities of the
thermodynamic quantities (specific heat, susceptibility, magnetization).

The structure of the paper is the following. Random quantum chain models studied in this
paper are presented in Sec.\ref{sec:random}. In Sec.\ref{sec:density}
the density of states of the low energy excitations is calculated by the Dyson-Schmidt
technique and the relation of this technique with the SDRG method is discussed.
Thermodynamic singularities are
calculated in Sec.\ref{sec:therm}
and the results are discussed in Sec.\ref{sec:disc}.

\section{Random quantum chains}
\label{sec:random}

\subsection{Random tight-binding model}

The first model we consider is a one-dimensional tight-binding model with off-diagonal
disorder\cite{cohen} being
defined by the Hamiltonian:
\be
{\cal H}=\sum_{i} t_i (|i\rangle\langle i+1|+|i+1\rangle\langle i|)\;,
\label{H_tb}
\ee
with random hopping matrix-elements, $t_i$.
The hopping matrix-elements are generally taken from different distributions at even ($t_e$) and odd ($t_o$)
sites, so that a quantum control-parameter is defined as:
\be
\delta=\frac{[\ln t_o]_{\rm av}-[\ln t_e]_{\rm av}}{{\rm var}(\ln t_e)+{\rm var}(\ln t_o)}\;,
\label{delta}
\ee
where $[\dots]_{\rm av}$ stands for averaging over quenched disorder and ${\rm var}(x)$
stands for the variance of $x$. For $\delta > 0$ ($\delta < 0$)
the model is asymmetric and the particles are preferentially at odd (even) bonds.
The symmetric model with $\delta = 0$ corresponds to a
quantum critical point.

In the basis, $|i\rangle$, the Hamiltonian is
represented by a tridiagonal matrix
\be
{\cal M}=
\begin{pmatrix}
0      &  t_1 &      &       &        &        \cr
 t_1 & 0      & t_2 &       &     &                   \cr
       & t_2  & 0    & t_3&     &                    \cr
       &        &   \ddots  &\ddots &\ddots   &                \cr
%       &        &      &     &  &  \lambda_{L-1}      &       \cr
%       &        &         & \lambda_{L-1}   & 0       &   h_L    \cr
% &       &   &          & h_L &  0             \cr
\end{pmatrix}
\label{M_tb}
\ee
and we are interested in its eigenvalue problem:
\be
{\cal M}\vec{\alpha}=E\vec{\alpha}
\label{M}
\ee
and the corresponding density of states, $n(E)$, at the center of the band.

\subsection{Random antiferromagnetic XX-chain}

The second model is the random antiferromagnetic XX-chain defined by the Hamiltonian:
\be
{\cal H}_{XX}=\sum_{i} J_i (S_i^x S_{i+1}^x+S_i^y S_{i+1}^y)
\label{H_XX}
\ee
in terms of the spin-$1/2$ operators, $S_i^{x,y}$, at cite $i$. Here the
$J_i$ exchange couplings are random variables which have different
distributions at even ($J_e$) and odd ($J_o$) sites. 
Using the Jordan-Wigner transformation:
$a_j^{\pm}=S_j^x \pm iS_j^y$ and $c^+_i=a_i^+\exp\left[\pi i \sum_{j}^{i-1}a_j^+a_j^-\right]$ and
$c_i=\exp\left[\pi i \sum_{j}^{i-1}a_j^+a_j^-\right]a_i^-$ this Hamiltonian is expressed in terms
of fermion creation ($c^+_i$) and annihilation ($c_i$) operators as\cite{lsm}:
\be
{\cal H}_{XX}=\sum_{i} \frac{1}{2}(J_i c^+_i c_{i+1} + {\rm h.c.})\;.
\label{ferm_XX}
\ee
The low-energy states of the model contain one fermion, which can be
written in the form $|\psi\rangle=\sum_i \alpha_i c^+_i |0 \rangle$, where
$|0 \rangle$ denotes the fermionic vacuum. Energies in this one fermion subspace
are obtained by the solution of
the eigenvalue problem of ${\cal M}$ in Eq.(\ref{M_tb})
with the correspondence:
\be
t_i=J_i/2
\label{corresp_XX}
\ee
Then the quantum control parameter of the model is just given by $\delta$ in Eq.(\ref{delta}).
In the asymmetric model with $\delta>0$ ($\delta<0$) there is enforced dimerization and
the system is in the random dimer phase\cite{HYMAN} with preference of odd (even) bonds. On the
other hand at the quantum critical point with $\delta=0$ the system is in the so called random
singlet phase\cite{fisherxx}.

\subsection{Random transverse-field Ising chain}

Our third and final model is the RTFIC, which is a prototypical model of random quantum systems having an order-disorder transition\cite{fisher}. This system is defined by the Hamiltonian:
\be
{\cal H}_{I} =
-\frac{1}{2}\sum_{i} \lambda_{i}\sigma_i^x \sigma_{i+1}^x-\frac{1}{2}\sum_{i=1}^L h_i \sigma_i^z
\label{eq:H}
\ee
in terms of the Pauli-matrices, $\sigma_i^{x,z}$, at site $i$ and the $\lambda_i$ couplings
and the $h_i$ transverse fields are random numbers.

As for the XX-chain ${\cal H}_{I}$ is expressed in terms of fermion operators\cite{pfeuty}:
\beqn
{\cal H}_{I}&=&
-\sum_{i}h_i\left( c^+_i c_i-\frac{1}{2} \right)\cr &-&
\frac{1}{2}\sum_{i}\lambda_i(c^+_i-c_i)(c^+_{i+1}+c_{i+1})
\label{ferm_I}
\eeqn
which is than diagonalized through a canonical transformation.
Now the low-energy excitations contain one free fermion, the possible energy of which is
given by the positive eigenvalues of the following symmetric matrix\cite{it,bigpaper}:
\be
{\cal T}=
\begin{pmatrix}
0      &  h_1 &      &       &        &                   \cr
 h_1 & 0      &\lambda_1 &       &     &                   \cr
       & \lambda_1  & 0    & h_2&     &                    \cr
       &        &   \ddots  &\ddots &\ddots   &                \cr
%       &        &      &     &  &  \lambda_{L-1}      &       \cr
%      &        &         & \lambda_{L-1}   & 0       &   h_L    \cr
%-w\lambda_L &       &   &          & h_L &  0             \cr
\end{pmatrix}
\label{T}
\ee
This is equivalent to ${\cal M}$ in Eq.(\ref{M_tb})
with the correspondences:
\be
t_{2i-1}=h_i,\quad t_{2i}=\lambda_i
\label{corresp_I}
\ee
Using this relation together with Eq.(\ref{delta}) the control parameter of the RTFIC is
given by the difference in the average log-fields and the average log-couplings.
For $\delta>0$ ($\delta<0$) the system is in the paramagnetic (ferromagnetic) phase,
and $\delta=0$ represents the quantum critical point.

We can thus conclude that the low-energy properties of all the three models are related to
the eigenvalue problem of ${\cal M}$ in Eq.(\ref{M_tb}).
In the next section we calculate the density of states
of matrix ${\cal M}$ around $E=0$ by the Dyson-Schmidt method.

\section{Density of states at the center of the spectrum}
\label{sec:density}
Here in the first two subsections we recapitulate the basic ingredients of the Dyson-Schmidt method
and present the solution in the continuum approximation. Our findings, which are obtained in the
strongly disordered regimes are presented in the last two subsections.
\subsection{The random walk method}
In order to calculate the density of states of ${\cal M}$ we introduce a new vector, ${\vec \Delta}$,
with the components $\Delta_i=\alpha_{i-1}t_{i-1}/\alpha_i$, which satisfy the equations : $\Delta_{i+1}=t_i^2/(E-\Delta_i)$. The basic ingredient of the Dyson-Schmidt method\cite{dyson}
is the {\it node counting theorem} of one-dimensional Hamiltonians, which states
that the integrated density of states, $N(E)=\int_{-\infty}^E n(E') {\rm d}E'$,
is given by the fraction of
positive terms in the sequence of $\Delta_i$. At the center of the band, $E=0$,
the components of ${\vec \Delta}$ have alternating signs, thus here the "sign variables" $s_i \equiv sign[\Delta_i (-1)^i]$
have a fully ordered state, $\dots \uparrow \uparrow \uparrow \uparrow \uparrow \uparrow \uparrow \uparrow \uparrow \uparrow \dots$ and $N(0)=1/2$. For nonzero $E$ the iterated equations
for ${\Delta}_i$ are the following\cite{eggarter_riedinger}:
\beqn
\Delta_{2i}&=& f_{2i-2} \left(\frac{t_{2i-1}}{t_{2i-2}}\right)^2\Delta_{2i-2} \cr
f_{2i-2}&=&\frac{1-E/\Delta_{2i-2}}{1+(E \Delta_{2i-2}-E^2)/t_{2i-2}^2}\;,
\label{iterated}
\eeqn
which lead to different iteration behaviors for small positive $E$ for various
limiting values of ${\Delta}_{2i}$. These are summarized as:
\begin{subequations}\label{limits}
\begin{align}
&\Delta_{2i+1}/\Delta_{2i}<0,& {\rm if}&  &\Delta&_{2i} < E \label{a} \\
&f_{2i}=1,& {\rm if}& &E& \gg \Delta_{2i} \gg {\tilde t}^2/E \label{b} \\
&\Delta_{2i+2}/\Delta_{2i} < 1,& {\rm if}& &\Delta&_{2i} \approx {\tilde t}^2/E \label{c}
\end{align}
\end{subequations}
where $\tilde{t}$ denotes the typical (average) value of the matrix-element. According
to (\ref{b}) we
can identify an interval, $[ E,{\tilde t}^2/E]$, in which the "signs" stays ordered, say $s_i=\uparrow$.
There is a finite upper boundary value at $\Delta_{max}={\tilde t}^2/E$,
where the iterated sequence is reflected, but $s_i$ stays $\uparrow$ (see (\ref{c}),
whereas as the sequence arrives at the lower boundary value, $\Delta_{min}=E$, the "spins"
change sign (see (\ref{a}) and the iteration process starts again, however in a new domain
with $s_i=\downarrow$. Consequently for a small $E>0$ the sign variables have a fragmented domain structure
$\dots \downarrow \downarrow \downarrow \downarrow \uparrow \uparrow \uparrow \uparrow \uparrow \downarrow \downarrow \downarrow \downarrow \dots$
and therefore  the fraction of positive terms in the sequence of $\Delta_i$ is somewhat larger
than $1/2$, due to extra positive terms appearing at the domain walls. If
the typical (average) size of a domain is
denoted by $\tilde{\ell}$, then the density of states is asymptotically given by:
\be
N(E)-N(0)=\frac{1}{2\tilde{\ell}}\;.
\ee
We can thus summarize that to obtain the density of states at the center of the
spectrum it is enough to follow the evaluation of the sequence, $\Delta_i$, within one
typical domain and calculate its size, $\tilde{\ell}$. Within this domain
we formally put $f_{2i-2}=1$ in Eq.(\ref{iterated}) and set i) a reflecting
boundary at $\Delta_{max}$ and ii) an absorbing boundary at $\Delta_{min}$.
If we introduce the logarithmic variable,
$\ln \Delta_{2i}=u_{2i}$, we obtain a random walk (directed polymer) problem:
\be
u_{2i}=2(\ln t_{2i-1}-\ln t_{2i-2})+u_{2i-2}\;,
\label{walk}
\ee
with reflecting $(u=u_{max})$ and absorbing $(u=u_{min})$ boundary conditions.
In this language the walker (polymer) starts at $u_0=u_{max}$ and
its mean first-passage time (length) at the position $u_{min}$ is just $\tilde{\ell}$, thus $u_{min}=u_{\tilde{\ell}}$.

\subsection{Analysis in terms of the diffusion equation}
\label{sec:diff}
In order to set the length-scales in the random walk problem we use a continuum approximation in which Eq.(\ref{walk}) is transformed into a diffusion equation:
\be
\frac{\partial P(u,\ell)}{\partial \ell}=D\frac{\partial^2 P(u,\ell)}{\partial u^2}
-v\frac{\partial P(u,\ell)}{\partial u}\;.
\label{diff}
\ee
Here $P(u,\ell)$ is the probability distribution of the walk, $D=2[var(\ln t_e)+var(\ln t_o)]$
is the diffusion coefficient and $v=2([\ln t_o]_{\rm av}-[\ln t_e]_{\rm av})$
is the drift velocity. The typical size of the transverse fluctuations of the walk is given by:
$\tilde{u}=D/v=\delta^{-1}$, whereas the average distance between two reflections, $\xi$, follows
from the relation $\tilde{u} \sim \sqrt{D\xi}$, thus we obtain for the correlation length:
\be
\xi \sim D^{-1} \delta^{-2}
\ee
which agrees with the result of SDRG calculations\cite{fisher}. The continuum approximation and thus the
use of the diffusion equation is justified
if the correlation length is much larger than the lattice spacing. This condition is
satisfied if we are either at the
critical point, $\delta=0$, or in the weakly disordered Griffiths phase
with $|\delta| \ll 1$.

\subsubsection{Critical point}

At the critical point both the correlation length, $\xi$, and the typical size of transverse
fluctuations, $\tilde{u}$, are divergent and they are related with the length scale, $\tilde{\ell}$,
as: $\xi \sim \tilde{\ell}$ and ${\tilde u} \sim \sqrt{D \tilde{\ell}}$. Absorption of the walker
in this case is due to {\it typical fluctuations}, when ${\tilde u}$ grows to the order of
the width of the strip: ${\tilde u} \sim \Delta u=u_{max}-u_{min}=\ln(\tilde{t}^2/E^2)$.
From this follows:
\be
\tilde{\ell} \sim \frac{1}{D} \ln(\tilde{t}^2/E^2)^2\;
\ee
so that
\be
N(E)-N(0)\sim D\left[\ln(\tilde{t}^2/E^2)\right]^{-2}\;.
\ee
This is the classical result derived by Eggarter and Riedinger\cite{eggarter_riedinger}.
\subsubsection{Weakly disordered Griffiths phase}
\label{sec:weak_G}
In the weakly disordered Griffiths phase with $1 \ll \delta>0$ the walker is drifted towards the
reflecting boundary and both the correlation length and the typical size of the transverse
fluctuations are finite, but much larger than the lattice spacing, thus the continuum
approximation is valid. In this case $\tilde{u}$ is much smaller than the width of the strip,
its absorption takes place with a very small probability:
$p(\Delta u) \propto \exp(-\frac{\Delta u}{\tilde{ u}})$, thus it is a rare region
effect and due to {\it extreme fluctuations}.
Before having such a large fluctuation the walker is reflected several times and the typical number of
independent excursions is given by $ \tilde{\ell}/\xi$,
the value of which follows from extreme-value
statistics\cite{galambos}: $p(\Delta u) \tilde{\ell}/\xi=O(1)$. From this we have
\be
\tilde{\ell} \sim \xi \exp\left(\frac{v}{D}\ln(\tilde{t}^2/E^2)\right) \sim \left(\frac{\tilde{t}}{E}\right)^{1/z};
\label{n_z}
\ee
with 
\be
\frac{1}{z}=\frac{2v}{D}=2\frac{[\ln t_o]_{\rm av}-[\ln t_e]_{\rm av}}{[var(\ln t_e)+var(\ln t_o)]}=2\delta\;.
\label{1/z}
\ee
Here $z$ is just the dynamical exponent defined in Eq.(\ref{tau}).
In the weakly disordered Griffiths phase at the center of the band there is a power-law
singularity of the density of states:
\be
N(E)-N(0)\sim \left(\frac{\tilde{t}}{E}\right)^{-1/z}\;.
\label{N_E}
\ee
which is equivalent to the form in Eq.(\ref{n_E}).
This result for the random antiferromagnetic XX-chain has been presented in\cite{lamas}.

\subsection{Analysis in the strongly disordered Griffiths phase}
\label{sec:asymptotic}
In the strongly disordered Griffiths phase the correlation length is in the order of
the lattice spacings and the continuum approximation is not valid. In this case we use
discrete variables and denote the (nonlogarithmic) position of the walker at the $j$-th step of
the $k$-th independent excursion, which starts at $r(k)$, as $\Delta_{2j}^{(k)}$. Thus $r(k)/k = \xi$
for large $k$ and the normalized position is given by:
\be
\rho_{2j}^{(k)} \equiv \frac{\Delta_{2j}^{(k)}}{\Delta_{max}}=
\prod_{j'=1+r(k)}^{j+r(k)} \left(\frac{t_{2j'-1}}{t_{2j'-2}}\right)^2\;.
\ee
The condition of absorption is formulated as:
\be
\min_{k} \min_{1<j<\Delta r(k)} \rho_{2j}^{(k)}=\frac{\Delta_{min}}{\Delta_{max}}=\frac{\tilde{t}^2}{E^2}\;,
\ee
where $\Delta r(k)=r(k+1)-r(k)$, which can be replaced by $\Delta r=\infty$. Keeping in mind that
$\rho^{(k)}_{2j}$ is typically much larger than its minimum value we can estimate the
order of magnitude of the minimum as
\be
\min_{k} \min_{1<j<\infty} \rho_{2j}^{(k)}
\propto \min_{k} \left[ y^{(k)} \equiv \sum_j (\rho_{2j}^{(k)})^{-1} \right]^{-1}\;.
\label{min2}
\ee
Here $y^{(k)}$ is a Kesten variable\cite{kesten} for any $k$,
the distribution function of which for large arguments displays a singularity:
\beqn
p(y) \opsim_{y \to \infty} y^{-(1+\mu)}\;.
\label{p_asymp}
\eeqn
where the exponent, $\mu$, is given by the positive root of the equation:
\be
\left[\left(\frac{t_o^2}{t_e^2}\right)^{\mu}\right]_{\rm av}=1\;.
\label{mu}
\ee
(For a pedagogical introduction to the theory of Kesten variables see Appendix C of Ref.[\onlinecite{im}].)
In this way the typical number of excursions, $ \tilde{\ell}/\xi$, follows from extreme-value
statistics\cite{galambos}: $\tilde{\ell}/\xi \int_{y_{max}}^{\infty} p(y) {\rm d} y=1$, and
we obtain:
\be
\tilde{\ell} \sim \xi \left(\frac{t}{E}\right)^{2\mu}\;.
\ee
Comparing with Eq.(\ref{n_z}) we see that the dynamical exponent in the strongly disordered
Griffiths phase is given by:
\be
\frac{1}{z}=2\mu\;,
\label{z_mu}
\ee
which in the limit $\delta\ll 1$ gives back the result obtained in the weakly disordered Griffiths
phase\cite{ijl,i02} in Eq.(\ref{1/z}). Then
with the correspondence in Eq.(\ref{z_mu}) the density of states at the center of the band is
given in Eq.(\ref{N_E}).

\subsection{Relation with the strong disorder renormalization group method}

The density of states in the center of the spectrum of $\cal M$ can be analyzed
by the SDRG method\cite{im}, too, and here we outline this procedure. The first
step in this study is to arrange the matrix-elements, $t_i$, in descending order and use
the largest one, $\Omega={\rm max}_i \{t_{i}\}$, to set the energy scale in the system.
Let us denote the largest term by $t_j$, which connects sites $j$ and $j+1$ and
eliminate the two equations in the eigenvalue problem which contain $t_j$.
In second-order perturbational method, which is correct up to $O\left((t_{j-1}/t_j)^2\right)$ and
$O\left((t_{j+1}/t_j)^2\right)$ we have for
the effective matrix-element, $t'$, between the remaining sites, $j-1$ and $j+2$:
\be
t' \approx \frac{t_{j-1} t_{j+1}}{t_j}\;.
\ee
This new term has a length, $m'=m_{j-1}+m_{j}+m_{j+1}=3$, where the original matrix-elements
have unit lengths.

In the following steps we repeat the decimation transformation, during which the energy scale is reduced,
the lengths are increased and the distribution functions of the matrix-elements, $R_e(t_e,\Omega)$ and $R_o(t_o,\Omega)$, approach their fixed-point form. This type of RG equations have been
analytically solved both at the critical point\cite{fisher} and in the Griffiths phase\cite{ijl,i02}.
Here we summarize the known results for the Griffiths phase with $\delta>0$.

In the starting steps of the RG both $t_e$ and $t_o$ terms are decimated, but the transformation
in later steps become asymmetric. As the typical
lengths are growing beyond $m' \sim \xi$ almost exclusively the $t'_o$ terms are decimated and the
$t'_e$ terms become very small, such that at the fixed point, $\Omega \to \Omega^*=0$,
we have $t'_e/t'_o \to 0$. As a consequence the energy of the
low-energy excitations is simply $E \simeq t'_o$. At the fixed point the distribution of $t_o$
is given by\cite{ijl,i02}:
\be
R_o(t_o,\Omega)=\frac{2 \mu}{\Omega} \left(\frac{\Omega}{t_o}\right)^{1-2\mu}\;,
\ee
where $\mu$ is defined in Eq.(\ref{mu}). This is just equivalent to the distribution of the excitation
energies in Eq.(\ref{n_E}), with the dynamical exponent defined in Eq.(\ref{z_mu}).

Now to make a correspondence with the random walk method the starting RG steps which
lead to an effective $t'_e(k)(\gg t'_o(k))$ of length $m'(k) \sim \xi$ are equivalent
to an excursion (between two reflections) of the walk of size $\Delta r(k) \sim \xi$ and the minimal value of $\rho^{(k)}_{2j}$ for this excursion is just the renormalized value of $t'_o(k)$. The analogous
quantities in the two approaches are collected in Table \ref{table:1}.
\begin{table}
\caption{Analogous quantities in the random walk (RW) and in the SDRG methods \label{table:1}}
 \begin{tabular}{|c||c|c|c|}  \hline
  method & independent unit & length scale & energy scale \\ \hline
  RW & excursion & size of the excursion & ${\rm min}_j \rho^{(k)}_{2j}$ \\ \hline
  SDRG & cluster & size of the cluster & $t'_o(k)$ \\
\hline
  \end{tabular}
  \end{table}

\section{Thermodynamic singularities}
\label{sec:therm}
Here we consider the random tight-binding model with half filling, as well as the random antiferromagnetic XX-chain and the RTFIC and note that all these models are expressed in terms of free
fermions. The common form of the Hamiltonians is given by:
\be
{\cal H}_F=\sum_q E_q (\eta_q^+ \eta_q -1/2)
\ee
where $E_q$ denotes the $q$-th eigenvalue of ${\cal M}$ and $\eta_q^+$ ($\eta_q$) are fermion
creation (annihilation) operators. The ground-state energy per site of this system is given by:
\be
{\cal E}=-\frac{1}{2L} \sum_q E_q=-\frac{1}{2}\int_{E_{min}}^{E_{max}} n(E) E {\rm d} E\;
\label{energy_0}
\ee
and the free energy per site:
\beqn
&{\cal F}&=-\frac{T}{L} \sum_q \ln \left[2\cosh \left(\frac{E_q}{2T} \right) \right]=\cr
&-&T\left\{\ln2 + \int_{E_{min}}^{E_{max}} n(E) \ln\left[\cosh \left(\frac{E}{2T} \right) \right] {\rm d} E \right\}\;
\label{free_energy}
\eeqn
where $L$ is the length of the chain.
From the free energy we obtain the internal energy:
\be
{\cal E}(T)=-\frac{1}{2}\int_{E_{min}}^{E_{max}} n(E) E \tanh\left(\frac{E}{2T} \right){\rm d} E
\ee
and the specific heat:
\be
c_v(T)=\int_{E_{min}}^{E_{max}} n(E)\left(\frac{E}{2T} \right)^2 \cosh^{-2}\left(\frac{E}{2T} \right){\rm d} E.
\ee
Now using the form of the density of states at the center of the band we obtain for the
low temperature behavior:
\be
c_v(T) \propto {\cal A} T^{1/z} \int_{-\infty}^{\infty} \varepsilon^{1/z+1} \cosh^{-2}
\varepsilon\ {\rm d} \varepsilon.\;,
\label{cv1}
\ee 
in agreement with the scaling result in Eq.(\ref{chi_cv}). Note that the prefactor in Eq.(\ref{cv1}),
${\cal A}$, is proportional to $\xi^{-1} z^{-1}$, which means that in the weakly disordered Griffiths
phase we have: ${\cal A} \sim \delta^3,~\delta \ll 1$, in agreement with the
SDRG result\cite{fisher}.

Next we consider the random antiferromagnetic XX-chain for which in the Hamiltonian in Eq.(\ref{H_XX}) we
introduce a homogeneous ordering field: $H\sum_i S_i^z$. This term with fermionic
variables assumes the form: $H/2 \sum_i (c_i^+c_i-1/2)$, thus the eigenvalue matrix $\cal M$
contains also diagonal elements: ${\cal M}_{i,i}=H/4,~\forall i$,
and the eigenvalues are shifted by $E \to E+H/4$. The magnetization is obtained through differentiation:
\be
m(H,T)=-\frac{\partial {\cal F}}{\partial H} \sim \int_{-H/4}^{H/4} n(E) \tanh\left(\frac{E}{2T} \right){\rm d} E
\label{m_H_T}
\ee
where we have used the fact that the spectrum of ${\cal M}$ in Eq.(\ref{M_tb}) is symmetric to $E=0$.
At zero temperature $m(H,0)$ is singular for small $H$:
\be
m(H,0)\sim N(H/4)-N(-H/4) \sim H^{1/z}
\ee
as in Eq.(\ref{m_H}). Evaluating the integral in Eq.(\ref{m_H_T}) for small $H$ and $T$,
however with $H/T=O(1)$
we obtain for the low-temperature susceptibility:
\be
\chi(T) \sim T^{1/z-1}
\ee
which corresponds to the scaling result in Eq.(\ref{chi_cv}).

\section{Discussion}
\label{sec:disc}

In this paper we have studied Griffiths-McCoy singularities in random quantum
(tight-binding, XX and Ising) spin chains,
which can be represented in terms of free fermions. The main step of our investigation is
the calculation of the density of states of the low energy excitations, which excitations
are eigenvalues of a symmetric tridiagonal matrix with random
entries, however with an odd-even asymmetry.
This latter problem is solved exactly by the Dyson-Schmidt technique\cite{dyson,eggarter_riedinger}
for any value of the quantum control parameter, $\delta$. Previous
studies of this problem are restricted to the quantum
critical point\cite{eggarter_riedinger}, $\delta=0$, and to the weakly disordered
Griffiths phase\cite{lamas}, $\delta \ll 1$.

As we described in Sec.\ref{sec:diff}
in this problem there are three length
scales: the mean-first passage length, $\tilde{\ell}$,
the correlation length, $\xi$, and the lattice spacing, $a$. In the different regimes
of the quantum control parameter their relative magnitudes are summarized in Table~\ref{table:2}.
\begin{table}
\caption{Relation between the length-scales in different regimes of the quantum
control parameter. \label{table:2}}
 \begin{tabular}{|c||c|c|}  \hline
  critical point & $\delta=0$ & $\tilde{\ell} \sim \xi \gg a$ \\ \hline
  weakly dis. Griffiths & $|\delta| \ll 1$ & $\tilde{\ell} \gg \xi \gg a$ \\ \hline
  strongly dis. Griffiths & $|\delta|=O(1)$ & $\tilde{\ell} \gg \xi \sim a$ \\
\hline
  \end{tabular}
  \end{table}

In a finite system there is still another length scale given by the size of the system, $L$, and
the mean-first passage length can not exceed this value: $\tilde{\ell}\sim L$.
Consequently the lowest excitation energy is limited
to $E_1 \sim L^{-z}$. In this case one is interested in the distribution of the
scaling combination, $E_1 L^{z}$, which in the random walk method in
Sec.\ref{sec:asymptotic} is obtained from the statistics of extremes.
Here we recall that $E_1$ is just the minimum value of a set of $L/\xi$ independent
random numbers, each having the same parent distribution in a power-law form, see Eq.(\ref{p_asymp}).
Consequently the distribution of $\epsilon_1=aE_1 L^{z}$ in the large $L$ limit
follows the Fr\'echet distribution\cite{galambos}:
\be
\tilde{P}_1(\epsilon_1)=\frac{1}{z} \epsilon_1^{1/z-1} \exp(-\epsilon_1^{1/z})\;,
\label{frechet}
\ee
where $a$ is a nonuniversal constant which depends on the amplitude of the tail in Eq.(\ref{p_asymp}).
Here one can go on and consider the second eigenvalue, $E_2$, or more generally the
$q$-th smallest eigenvalue, $E_q$. These
are all obtained from the theory of extreme value statistics of independent and identically distributed
({\it i.i.d.})
random numbers and their distribution is given by the generalized Fr\'echet distribution,
see in\cite{galambos}. In this way
we have shown that the distribution of the lowest energy levels of these strongly correlated
physical systems are described in a form which holds for {\it i.i.d.} random
numbers. This scenario, which is shown here exactly for the specific models is expected to hold
generally for all such random quantum systems, even in higher dimensions, for which the low-energy behavior
is controlled by a so called strong disorder fixed point in the SDRG framework\cite{extr}.

The dynamical exponent, $z$, which is calculated exactly in this paper is found a continuous
function of the control parameter, $\delta$. Using the SDRG approach the same result is obtained\cite{ijl,i02},
thus our present study gives further credit to the conjecture that the SDRG method
provides asymptotically
exact results even far outside the critical point, as far as dynamical quantities are considered.
This latter statement is expected to hold for all systems with a strong disorder fixed point.

\begin{acknowledgments}
This work has been supported by the National Office of Research and
Technology under Grant No. ASEP1111 and by the Hungarian National Research Fund 
under grant No OTKA TO48721, K62588, MO45596.
\end{acknowledgments}

\end{document}